\documentclass[aps,prc,twocolumn,superscriptaddress,amsmath,showpacs,amssymb]{revtex4-1}
\usepackage{bm,physics}
\usepackage[T1]{fontenc}
\usepackage{textcomp}
\usepackage[dvips,final]{graphicx}
\usepackage{ulem, color}
\usepackage{hyperref}
\bmdefine{\ba}{a}
\bmdefine{\bb}{b}
\bmdefine{\bx}{x}
\bmdefine{\by}{y}
\bmdefine{\bz}{z}
\bmdefine{\bn}{n}
\bmdefine{\bp}{p}
\newcommand{\BM}{\begin{pmatrix}}
\newcommand{\EM}{\end{pmatrix}}

\begin{document}
\title {Evidence  of a higher nodal band $\alpha$+$^{44}$Ca cluster state  in   fusion reactions  and \\ $\alpha$ clustering      in $^{48}$Ti
 }

\author{S.~Ohkubo}
%\email{shigeo.ohkubo@rcnp.osaka-u.ac.jp}
\affiliation{Research Center for Nuclear Physics,
Osaka University, Ibaraki,
Osaka 567-0047, Japan}
\date{\today}
\begin{abstract}
\par
 In the nucleus    $^{48}$Ti,  whose structure  is essential in evaluating the half-life of the neutrinoless double-$\beta$ decay ($0\nu\beta\beta$) of $^{48}$Ca,
 the existence of  the  $\alpha$ cluster structure  is shown for the first time.
A unified description of scattering and structure is performed  for the $\alpha$  +$^{44}$Ca system.
By  using a global potential, which reproduces experimental  $\alpha$ +$^{44}$Ca scattering over a wide range of incident energies,  $E_\alpha$=18 - 100 MeV, it is shown that the observed $\alpha$ +$^{44}$Ca fusion excitation function at  $E_\alpha$=9 -18 MeV is  described well. The bump at $E_\alpha$=10.2 MeV is found to be due to a resonance which is  a $7^-$ state of the higher nodal band with the $\alpha$ +$^{44}$Ca cluster structure in  $^{48}$Ti. The local potential $\alpha$ +$^{44}$Ca cluster model locates the ground band of $^{48}$Ti in agreement with experiment and reproduces the enhanced $B(E2)$ values in the ground band  well. This shows that collectivity due to $\alpha$ clustering in $^{48}$Ti  should be taken into account  in the evaluation of the nuclear matrix element in the   $0\nu\beta\beta$ double-$\beta$ decay of $^{48}$Ca.
 \end{abstract}

\maketitle

\par
\section{INTRODUCTION}
 $\alpha$  clustering is essential not only in the light 
 $0p$-shell and $sd$ shell regions     \cite{Suppl1972,Suppl1980} but also   
in the medium-weight $fp$ shell region \cite{Ohkubo1998,Ohkubo1999} 
% in the medium-weight $fp$ shell region 
%\cite{Ohkubo1998A,Ohkubo1999}
as evidenced typically in the  $^{44}$Ti  region \cite{Michel1986,Michel1988,Ohkubo1988,Yamaya1990B,Guazzoni1993,%Ohkubo1998,%
Michel1998,Yamaya1998,Sakuda1998,Ohkubo1998B,Fukada2009}.
  Recent evidence    of the higher nodal band states with the 
$\alpha$ cluster structure in  $^{52}$Ti \cite{Ohkubo2020}, in which the intercluster relative motion is excited,      suggests that $\alpha$ clustering may persist in   nuclei  in-between such as    $^{48}$Ti  and $^{46}$Ti.
In fact,  the $\alpha$ spectroscopic factors 
 of $^{48}$Ti  and $^{46}$Ti   in the ($^6$Li,d)  $\alpha$-transfer reactions \cite{Anantaraman1975,Fulbright1977} are  larger than   that
 of $^{52}$Ti, which is the minimum   in the $A$=36-64 mass region  \cite{Anantaraman1975}. This is also  confirmed  in the  $(p,p\alpha$)  reactions  \cite{Carey1981,Carey1984,Roos1984}.
  Reference  \cite{Oertzen2001} reports that     the  excess neutrons outside the core 
      work as covalent bonding between the clusters. 

\par
 The structure of  $^{48}$Ti  is crucial  in determining the nature of neutrino, Dirac  or Majorana particle in the  measurements  of  neutrinoless double-$\beta$ decay $0\nu\beta\beta$  of $^{48}$Ca \cite{Tetsuno2020}.  The study of $0\nu\beta\beta$ \cite{Vergados2012},  which  violates lepton number conservation,  serves to solve the  longstanding fundamental questions beyond the standard model. 
 The inverse   half-life of $0\nu\beta\beta$  of $^{48}$Ca$(0^+)$ $\rightarrow$ $^{48}$Ti$(0^+)$ is given by  
$ [T^{0\nu}_{1/2}]^{-1}=G_{0\nu} \left|{<m_{\beta\beta}>/}{m_e}\right|^2 |M^{0\nu}|^2$,
where  $<m_{\beta\beta}>$   is the effective Majorana neutrino mass,
${m_e}$ is the electron mass, and $G_{0\nu}\sim 10^{-14}$ yr$^{-1}$ is a phase-space factor.  For the evaluation of the 
  nuclear matrix element  of the transition $M^{0\nu}$ \cite{Engel2017}, 
 it is essential to know the  ground state wave function of $^{48}$Ti accurately. 

Several  approaches  such as the shell model \cite{Caurier2008,Horoi2013,Iwata2016,Coraggio2020}, {ab initio} calculations \cite{Yao2020,Belley2021}, quasi-particle random phase approximation  \cite{Simkovic2013,Hyvarinen2015,Terasaki2015}, the projected Hartree-Fock Bogoliubov model \cite{Rath2010}, the generator coordinate method (GCM) \cite{Rodriguez2010,Hinohara2014}, the energy density functional \cite{Vaquero2013,Yao2015} and  the interacting boson model \cite{Barea2012} 
have been reported. 
As for the $\alpha$ clustering aspects of $^{48}$Ti,   a microscopic $\alpha$+$^{44}$Ca cluster model calculation in the GCM
  was performed \cite{Langanke1982,Wintgen1983}.
 However,  no state corresponding to the ground state of $^{48}$Ti was  obtained.
Experimentally in the recent   challenges \cite{Bailey2021,Bailey2019}, no $\alpha$ cluster states  have been observed in  $^{48}$Ti.
    A discovery   of  a typical $\alpha$ cluster state, such as       the higher nodal band states       observed  in   $^{52}$Ti         \cite{Ohkubo2020}, 
would shed light  on the $\alpha$ clustering of  the ground state of $^{48}$Ti.

  \par
 The purpose of this paper is to show   that the existence of a higher nodal band
  $\alpha$+$^{44}$Ca cluster structure  in $^{48}$Ti is evidenced   for the first time by  investigating the observed $\alpha$+$^{44}$Ca fusion cross sections  with a   global Luneburg lens-like  \cite{Michel2002,Ohkubo2016,Luneburg1964}
  potential that describes 
    $\alpha$ +$^{44}$Ca scattering over a wide range of incident energies,  $E_\alpha$=18 - 100 MeV.
 By using the  local potential $\alpha$+$^{44}$Ca cluster model 
 the observed  enhanced $B(E2)$ values of the ground band of $^{48}$Ti are reproduced well.
It is shown that the enhanced  $B(E2)$ values are caused by 
  $\alpha$ clustering    in $^{48}$Ti.
  
  The paper is organized as follows. Section II is devoted to the analysis of  $\alpha$+$^{44}$Ca scattering and  fusion reactions using a global potential. In Sec. III  $\alpha$+$^{44}$Ca cluster structure in $^{48}$Ti is studied. In Sec. IV   discussions and a summary are  given.  
  
\section{ANALYSIS OF $\alpha$+$^{44}$C\lowercase{a} SCATTERING  AND FUSION REACTIONS}
\par
In exploring the  $\alpha$ cluster structure  in medium-weight mass region where the level density is high, a unified  description of $\alpha$ scattering including rainbow scattering, prerainbows, backward angle anomaly (BAA) or anomalous large angle scattering  and the $\alpha$ cluster structure in the bound and quasi-bound energy region has been very powerful  \cite{Michel1986,Michel1988,Ohkubo1988,Michel1998,Ohkubo1999,Ohkubo2020}.  
 In fact, the $\alpha$ cluster structure in the $^{44}$Ti region was successfully explored from this viewpoint and   the predicted $\alpha$ cluster $K=0^-$ band with the $\alpha$+$^{40}$Ca cluster structure  \cite{Michel1986,Michel1988,Ohkubo1988}, which is a parity-doublet partner of the ground band of   $^{44}$Ti, was observed in experiment \cite{Yamaya1990B,Guazzoni1993}.  Systematic theoretical and experimental studies in the $^{44}$Ti region \cite{Ohkubo1998,Michel1998,Yamaya1998,Sakuda1998,Ohkubo1998B,Fukada2009}  confirmed  the existence of the  $\alpha$ cluster  in the beginning of the $fp$-shell.

\par

% Fig. 1     18 MeV   alpha+44Ca  scattering 
\begin{figure}[t!]
\includegraphics[width=8.6cm]{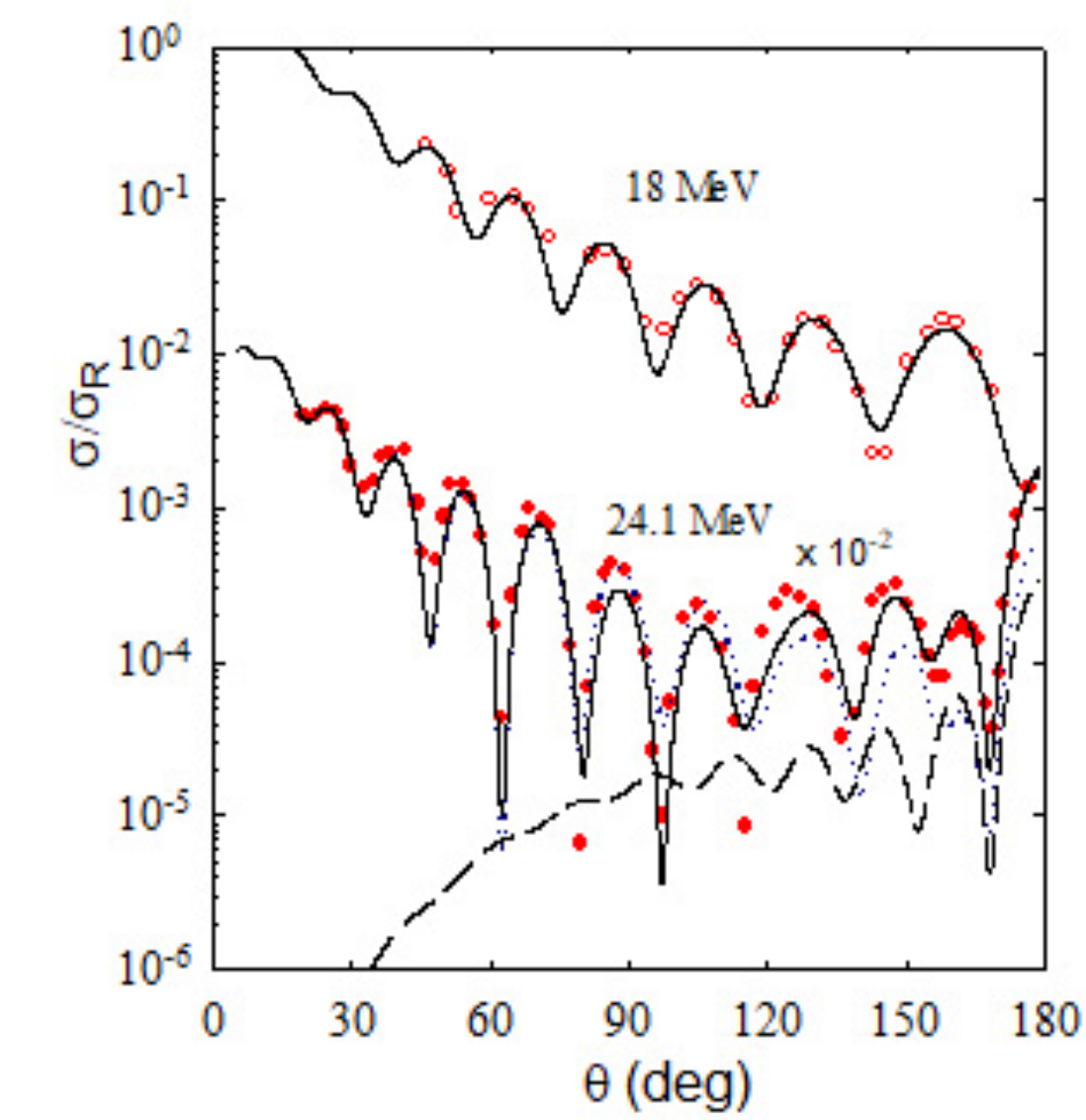}% Here is 
 \protect\caption{(Color online) Angular distributions 
  in   $\alpha$+$^{44}$Ca scattering at $E_\alpha$=18  and 24.1 MeV calculated with the optical  potentials in Table I (solid lines) are compared with the experimental data (circles)  \cite{Gaul1969}. The  calculated  angular distributions    are decomposed into the  barrier-wave (dotted lines) and the internal-wave (dashed lines) contributions. At 18 MeV the solid line overlaps with the dotted line and the dashed line is negligibly small and is not drawn. 
 }
\end{figure}

The  BAA in $\alpha$ particle scattering,  which was first observed for $\alpha$+$^{40}$Ca \cite{Budzanowski1966,Gruhn1966},  was systematically investigated in comparison with the isotopes  $^{42}$Ca and $^{44}$Ca \cite{Gaul1969} at low energies, $E_\alpha$=18-29 MeV.  It was concluded  \cite{Gaul1969}  that $\alpha$+ $^{44}$Ca scattering with no backward enhancement, which   is described well by the so-called standard optical with a Wood-Saxon form factor, is normal.  The normal behavior of the  angular distributions was attributed to strong absorption  due to the excess  neutrons  outside the $^{40}$Ca core \cite{Gaul1969,Delbar1978,Langanke1982,Wintgen1983}
 and little attention has been  paid to the $\alpha$+$^{44}$Ca cluster structure of  $^{48}$Ti in contrast to the $\alpha$+$^{40}$Ca cluster structure in $^{44}$Ti.

%%     Table I potential parameters 
\begin{table}[t!]
\begin{center}
\protect\caption{The optical potential parameters used in Fig.~1 and the volume integrals per nucleon pair, $J_v$, in unit of MeVfm$^3$ for the real potentials. 
$E_\alpha$, $V$, $W$ and $W_s$ are in units of MeV and $r_v$, $a_v$, $r_w$, $a_w$, $r_s$ and  $a_s$ are in units of femtomeers.}
\begin{tabular}{ccccccccccc}
 \hline
  \hline
  $E_\alpha$  &    $J_v$      &   $V$ & $r_v$ & $a_v$ &     $W$ &$r_w$& $a_w$& $W_s$& $r_s$ & $a_s$  \\
  \hline
  18   &     388    &  181 & 1.42 & 1.25    &  14.0 &1.75 &0.934 
       &   53.2&1.36&0.378\\
   24.1   &    356    &  166 & 1.42 & 1.25    &  10.6 &1.75 &0.934 
       &   17.0&1.36&0.378       \\    

 \hline
\end{tabular}
\end{center}
\end{table}

I extend  the global optical potential determined in $\alpha$+$^{44}$Ca scattering  over a wide range of energies of  $E_{\alpha}$=24 -100 MeV 
 \cite{Delbar1978}  to the lower energies $E_{\alpha}$=9 - 24 MeV.
The optical   potentials are given by
   $U(r) = - V f^2(r;R_v,a_v)+ V_{Coul}(r) -i W f^2(r;R_w,a_w) -i 4a_s W_s\frac{d}{dr}f^2(r,R_s,a_s)$
 with $f(r;R_i,a_i) =1/\{1+\exp[(r-R_i)/a_i]\}$. The Coulomb potential is assumed to be a uniformly charged sphere with a reduced radius  $r_{c}$=1.3 fm.  
First,  I show that   the experimental angular distribution in $\alpha$+$^{44}$Ca scattering at the lowest  energy  $E_\alpha$=18 MeV is reproduced well by the global potential. In Fig.~1 the calculated angular distributions at 18 MeV 
and 24.1 MeV  are compared with the experimental data. The potential parameters  $V$, $W$ and $W_s$ are searched with  other parameters  fixed as  in Ref. \cite{Delbar1978}, which are listed in Table~I. 
  The fits to the experimental data are much better than the ones calculated with the standard optical model in Ref. \cite{Gaul1969} and the improvements  are entirely  due to the Luneburg lens-like shape of the real potential \cite{Michel2002,Ohkubo2016,Luneburg1964}.
 To see why  BAA of the cross sections   is absent, the calculated angular distributions  are decomposed using the technique of Ref.\cite{Albinski1982} into the barrier-wave component reflected at the surface and the internal-wave component, which penetrates deep  into the internal region of the potential \cite{Brink1985}.
In Fig.~1, the barrier-waves dominate and the internal-waves hardly contribute, not visible at 18 MeV. 
The absence of the BAA    in  $\alpha$+$^{44}$Ca scattering  is 
  due to  no internal-wave contributions under  strong absorption,
   and not due to the real part of the potential since it is    similar to the  $\alpha$+$^{40}$Ca system as will be shown in Fig.~6. 
 
\par
Second,   I investigate the  $\alpha$+$^{44}$Ca fusion cross sections below  $E_\alpha$=18 MeV. 
In contrast to $\alpha$+$^{40}$Ca 
\cite{Eberhard1979,Michel1986A, Michel1986,Ohkubo1987A}, the  $\alpha$+$^{44}$Ca fusion excitation function  with no pronounced oscillations
 \cite{Eberhard1979} has never been paid attention to in the past decades.
 However,  one note in Fig.~2 that  the observed excitation function is not  monotonic     with  a bump at around  $E_\alpha$=10.2 MeV where absorption is relatively small near the threshold. The origin of  this bump, which   may be a  remnant of  the fusion oscillations    seen   typically   for $\alpha$+$^{40}$Ca \cite{Eberhard1979,Michel1986A,Ohkubo1987A}, $^{12}$C+$^{12}$C \cite{Ohkubo1987B},  $^{16}$O+$^{12}$C and $^{16}$O+$^{16}$O \cite{Mosel1984}, seems to be  related to the  $\alpha$+$^{44}$Ca molecular resonance.

%         alpha+44Ca   calculated 
% Fig.2 fusion cross sections in  comparison with the experimental data 
\begin{figure}[t!]
\includegraphics[width=7.6cm]{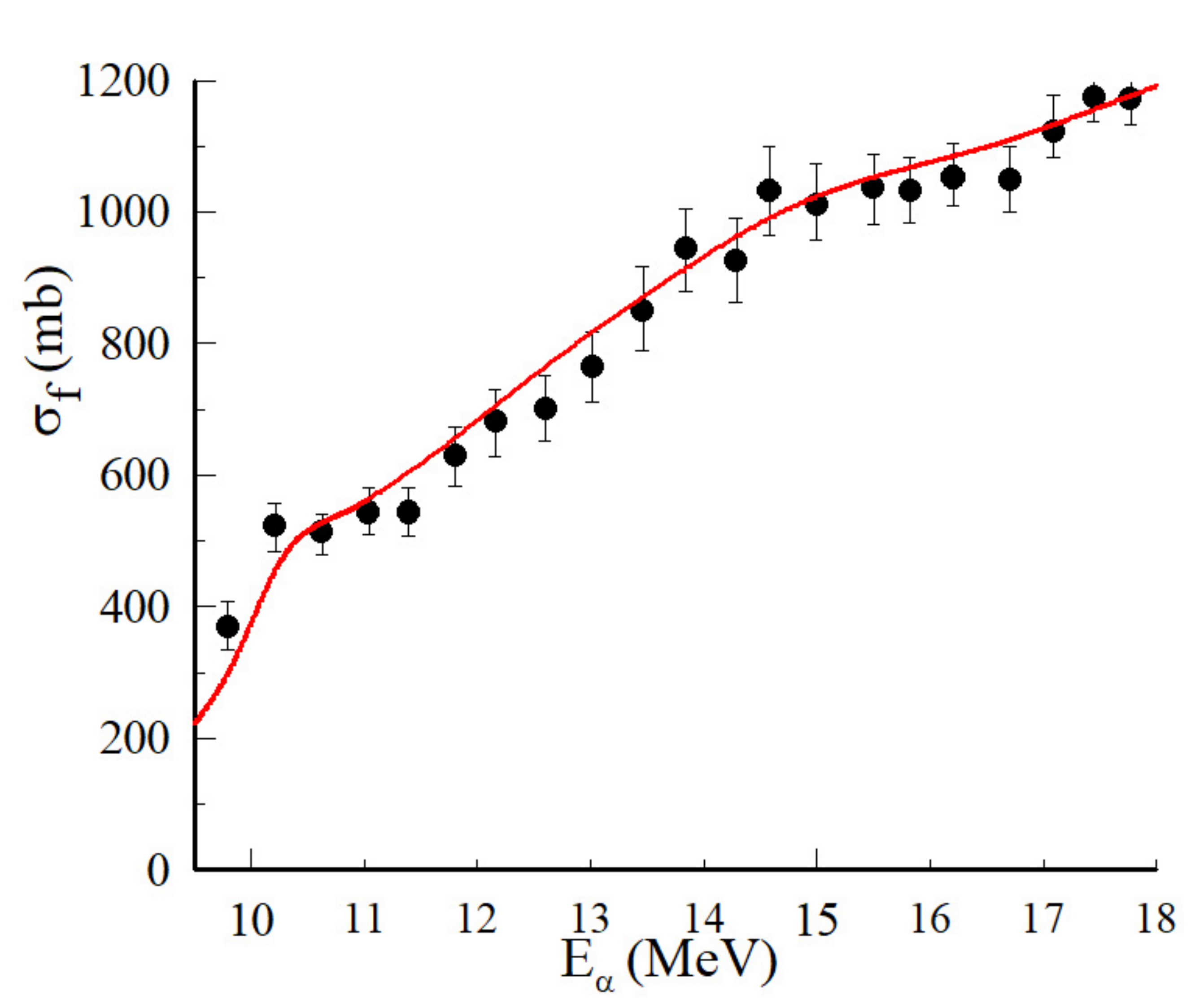}% Here is how to import EPS art
 \protect\caption {(Color online) Calculated   excitation function of the  $\alpha$+ $^{44}$Ca fusion cross sections (solid lines)
  is compared with  the experimental data (filled circles) \cite{Eberhard1979}.
  }
\end{figure}

 \par
 In calculating the  fusion cross sections using the optical potential, I followed the prescription used for $\alpha$+$^{40}$Ca in Ref.\cite{Michel1986A} where the fusion cross section is defined as a total reaction cross section due to the short-ranged imaginary potential.  The fusion  cross sections  at $E_\alpha$< 18 MeV are calculated using   the optical potential  at $E_\alpha$=18 MeV  in Table I with $r_w$=1.3 for the volume imaginary potential and $W_s$=0 for the surface imaginary potential due to direct reactions.  
   The  energy dependence of the strength parameter, $W$, which was originally determined  at $E_\alpha$>24 MeV in Ref. \cite{Delbar1978}
   is modified for  $E_\alpha < $18 MeV to decrease  linearly toward the barrier top energy due to the dispersion relation of the  threshold anomaly \cite{Mahaux1986}, 
$W$=$ a E_\alpha+b$ for   $E_\alpha \geq8.4$ MeV with      $a$=1.281 and $b$=-10.74.

  \par
 In Fig.~2 the  calculated  fusion excitation function 
  is compared with the experimental data \cite{Eberhard1979}.  
It is to be noted that the  bump at $E_\alpha$=10.2 MeV, which was not reproduced in Ref. \cite{Eberhard1979}, is reproduced well by the calculation. Also the calculated fusion cross sections do not decrease monotonically  toward the threshold energy in accordance with    the very smooth oscillatory behavior of the experimental data, which is a remnant of the pronounced oscillatory structure seen 
in the fusion excitation function for  $\alpha$+$^{40}$Ca 
  in the same energy region. 
  The emergence of a bump at  $E_\alpha$=10.2 MeV is due to the relatively weak absorption  near the barrier top energy,    $E$=6.2 MeV, which corresponds to $E_\alpha$=7.2 MeV.

 % Fig.3 fusion partial cross sections  
\begin{figure}[t!]
\includegraphics[width=7.6cm]{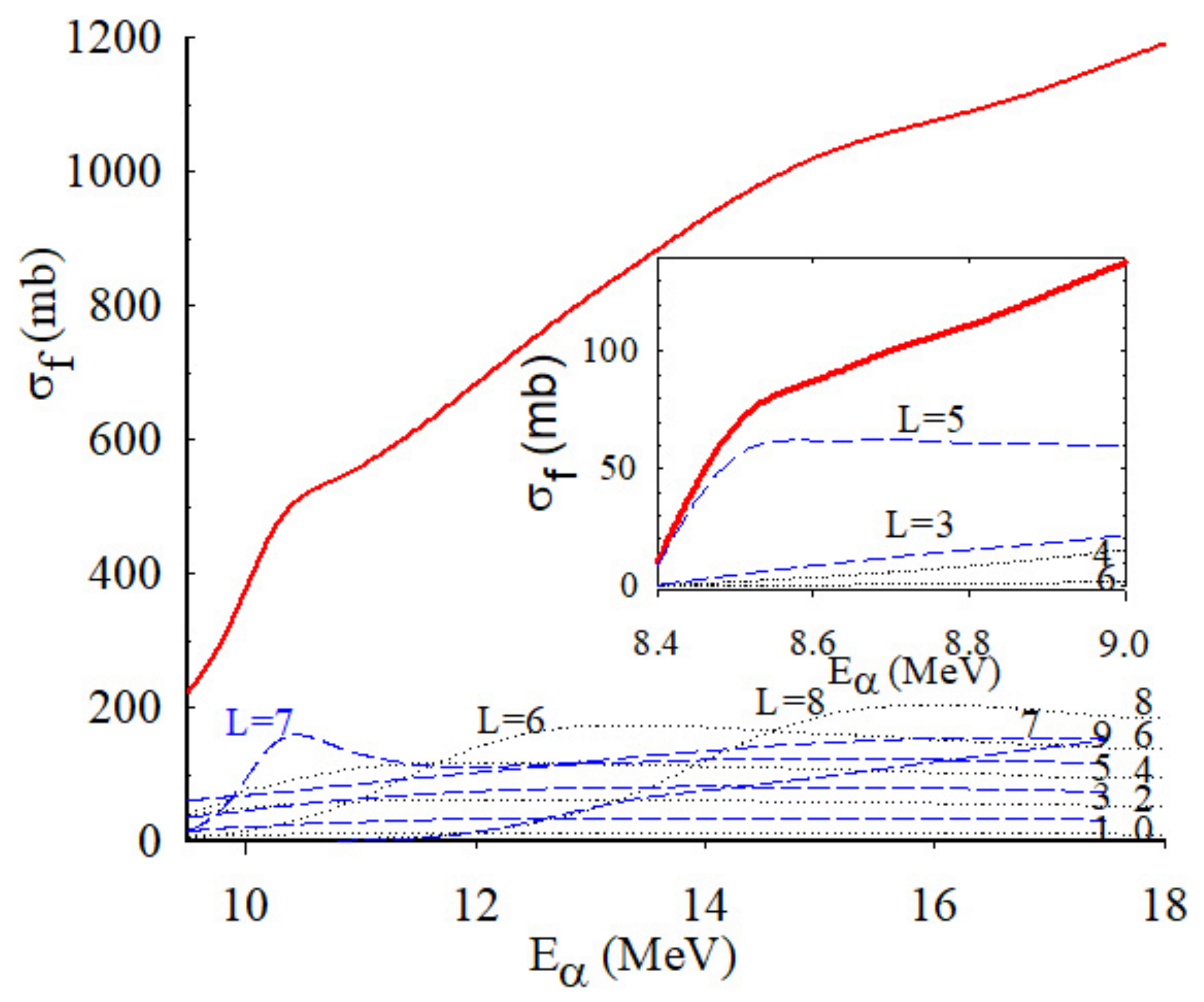}% Here is how to 
 \protect\caption {(Color online) Calculated  excitation functions of the  $\alpha$+ $^{44}$Ca  fusion cross sections  (solid lines)  are decomposed into the partial wave contributions; even $L$ (dotted lines) and odd $L$ (dashed lines).  The inset is for   $E_\alpha$=8.4-9.0 MeV.
   }
\end{figure}
 \par
 In Fig.~3 the calculated fusion cross sections are decomposed into the partial cross sections. One notes that the bump peak is caused by the partial fusion cross sections with $L$=7 (dashed line). The broad resonance-like  behavior of the $L$=6 and 8 partial fusion cross sections contributes to  the non-monotonic behavior of the fusion excitation function.  The calculated  fusion excitation function   at  $E_\alpha$=8.4- 9 MeV  is displayed in the inset.  Although the magnitude of the cross sections becomes  smaller, one notices the appearance of  a  bump at  $E_\alpha$=8.5-8.6  MeV  where the absorption is much  smaller than the bump at $E_\alpha$=10.2  MeV.  It is found that this structure is caused by the $L=$5 partial cross sections (dashed line). 

 \par
In order to reveal the origin of the bump structures in the fusion excitation function, 
in Fig.~4  the phase  shifts calculated with  $V=$181 MeV  potential (referred to as D181 hereafter)  at  $E_\alpha$=18 MeV by switching off the imaginary potentials are displayed. One notes that the resonance at   $E_\alpha$=10.3 MeV with a width $\Gamma_{c.m.}$=0.18 MeV is responsible for the peak of the partial fusion cross section for $L=$7 and therefore for the bump observed at around $E_\alpha$=10.2 MeV. The resonance at $E_\alpha$=8.3 MeV with  $\Gamma_{c.m.}$=0.22 MeV is also responsible for the bump  at  $E_\alpha$=8.3 MeV in the inset of Fig.~3. There appear resonances for $L=$9 and $11$ at  $E_\alpha$=12.96  ($\Gamma_{c.m.}$=0.11 MeV)  and 16.07  ($\Gamma_{c.m.}$=0.04 MeV), respectively, however, the contributions to the fusion cross section are not seen clearly.
 The broad $L=$6 and 8 resonances  at around   $E_\alpha$=12 MeV and 16 MeV, respectively,  are  also the origin of  the non-monotonic behavior of the
  fusion excitation function.

% Fig.4  D181 W=0 phase shift   
\begin{figure}[t!]
\includegraphics[width=7.6cm]{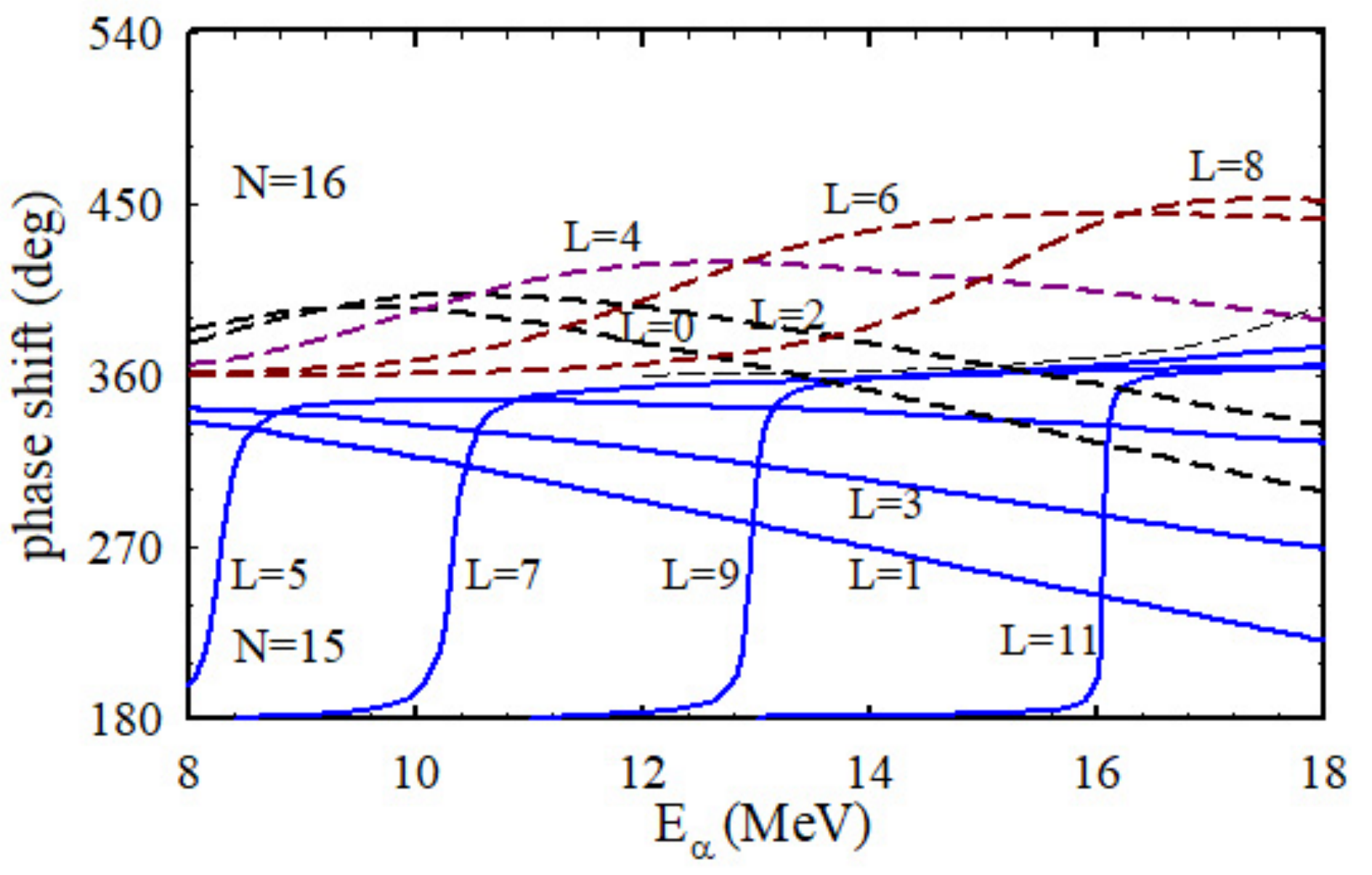}% Here is how to import 
 \protect\caption {(Color online) The phase shifts  for $\alpha$+ $^{44}$Ca scattering calculated 
  with the D181 real  potential. The vertical axis is  $\delta_L$-$m_L$$\pi$ where $m_L$ is the number of the Pauli-forbidden states. See the  text. The resonances  for odd $L$ and  even $L$ correspond to   $N=15$ and $N=$16, respectively. 
  }
\end{figure}

\section{ $\alpha$+$^{44}$C\lowercase{a} CLUSTER STRUCTURE IN $^{48}$T\lowercase{i}}
  \par
How these resonances are  understood  as the highly excited states  with the  
 $\alpha$+$^{44}$Ca cluster structure in $^{48}$Ti?
 In Fig.~5(a) the  energy levels calculated  with the D181 potential is displayed. The number of the Pauli-forbidden redundant states in the $\alpha$+$^{44}$Ca cluster model  \cite{Langanke1982} is  $m_L=(12 - L)/2$ for even $L \leq 12$ and   $m_L=(13 - L)/2$ for odd $L\leq13$. $m_L=0$  for $L\geq14$. 
According  to the generalized Levinson theorem \cite{Swan1955}, the phase shift $\delta_L$ in which the existence of the Pauli-forbidden states are taken into account  should satisfy     $\delta_L$ =$m_L \pi$ at $E_\alpha$=0  and  tends asymptotically to    $\delta_L$ =0 at $E_\alpha$=$\infty$. Shown in Fig.~4  are the phase shifts, $\delta_L$-$m_L$$\pi$, to make easy to see  the band structure of $N=15$ and $N=16$ where  $N= 2n$+$L$ with    $n$ being    the number of nodes in the  relative wave function   with the $\alpha$+$^{44}$Ca molecular structure. 
 Surprisingly the calculated  lowest Pauli-allowed $N=$12 band, which satisfies the Wildermuth condition due to the Pauli principle,
   falls   in  correspondence  to the experimental ground band of $^{48}$Ti.
One finds that the $L=7$ resonance that contributes to the bump of the fusion cross section at   $E_\alpha$=10.2 MeV is a member state of the higher nodal  $N=15$ band with  the   $\alpha$+$^{44}$Ca cluster structure, in which the intercluster relative motion is one more excited compared with the $N=13$ band.
This $7^-$ of $N=15$ is a second example of the higher nodal band with  negative parity in addition to  $^{44}$Ti \cite{Ohkubo1998B,Michel1998},
which  gives strong support to the  persistence of the  $\alpha$ cluster structures in $^{48}$Ti.
 It is highly desired to observe the   $5^-$ state  theoretically predicted 
  at  $E_\alpha$=8.3 MeV as well as the   $3^-$ and  $1^-$ states  of the  $N=15$ band  in a precise   experiment   such as sub-barrier fusion reactions    \cite{Beck2020,Jiang2021} and transfer reactions  etc.,
   although the $\alpha$-strengths as well as  the  $N=13$ and  $14$  bands, may be fragmented as  in $^{40}$Ca and $^{44}$Ti \cite{Yamaya1998,Sakuda1998,Michel1998,Ohkubo1998B}.

 \par
 The reason why the 
  $7^-$ state  of $N=15$ is observed in Fig.~2
  is related to  the considerably  high   $\alpha$ threshold energy of  $^{48}$Ti, 9.45 MeV  due to a non- $\alpha$ nucleus.
  Because of this, although  the excitation energy $E_x$=18.8 MeV of the $7^-$ state is  high, the energy from the   $\alpha$ threshold becomes  relatively small.
 The $N=14$ band, which is a higher nodal band of the ground band $N=12$, starts just above the $\alpha$ threshold. 
Its  observation in addition to the known analog higher nodal bands
 in   $^{20}$Ne \cite{Nemoto1972,Hiura1972,Ohkubo1977,Fujiwara1980},
 $^{40}$Ca \cite{Yamaya1998,Sakuda1998},  $^{44}$Ti \cite{Ohkubo1998B,Michel1998,Yamaya1998} and 
$^{52}$Ti \cite{Ohkubo2020,Bailey2019} would also reinforce the $\alpha$ cluster structure in  $^{48}$Ti. 
 
%%% Fig.5   Energy level 48Ti
 \begin{figure}[t!]
\includegraphics[width=8.6cm]{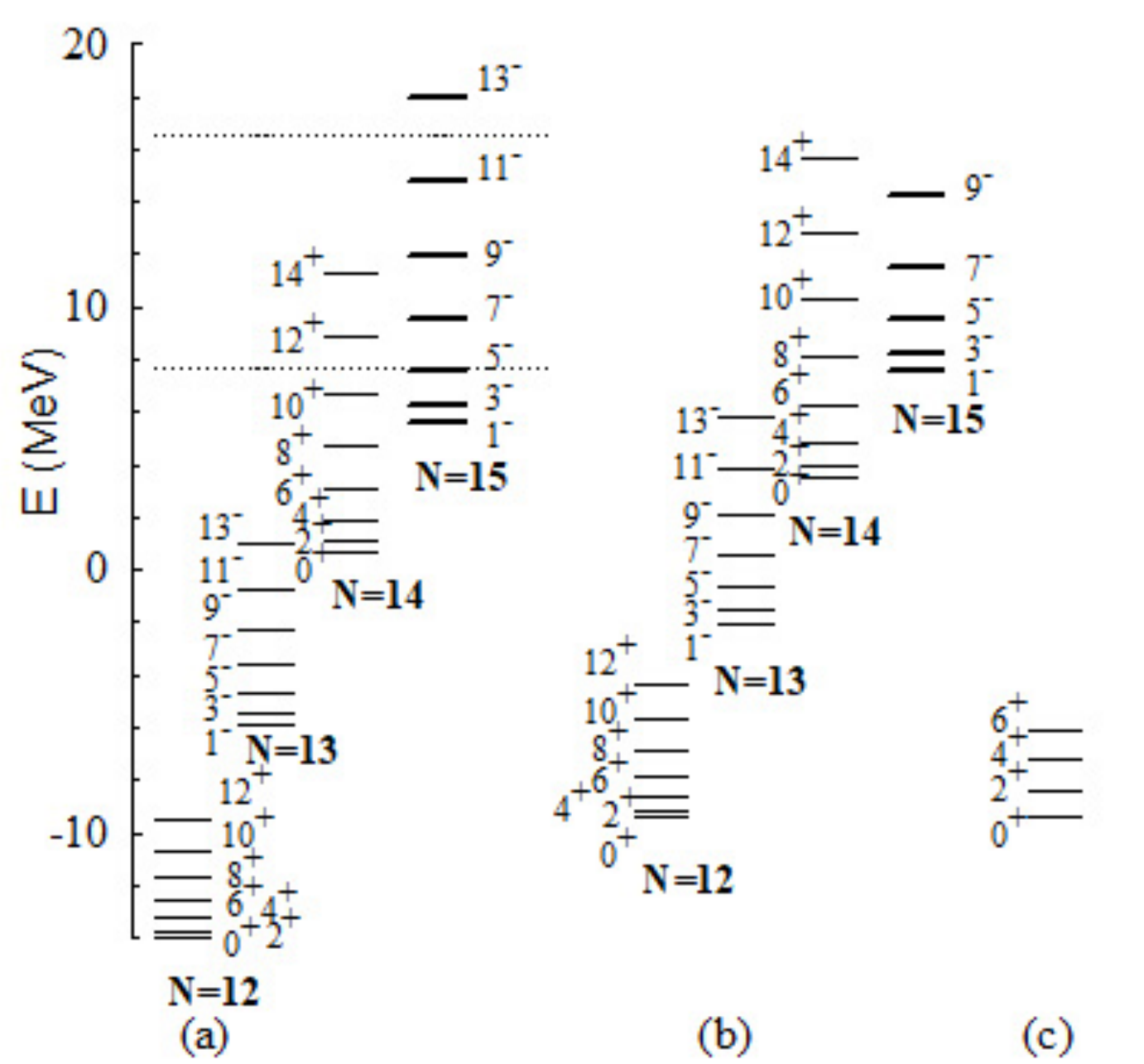}% Here is how to import 
 \protect\caption{ Energy levels of $^{48}$Ti with respect to the $\alpha$ threshold calculated with    the potentials      (a)  $V$=181 MeV and    (b) $V$=171 MeV    are compared  with  (c) the  experimental ground state band  \cite{Burrows2006}.    The energy levels are calculated in the bound state approximation except the $N$=15 band resonances. The      horizontal dotted lines indicate   the incident energy $E_\alpha$=8.4 - 18 MeV      of the fusion excitation functions in Fig.~3. 
 }
 \end{figure}

% Table II  excitation energies removed 2021.7.23
\begin{table}[t!]
\protect\caption{  Calculated intercluster rms radii (femtometers)
and  $B(E2)$ values (W.u.) for  the $J\rightarrow (J-2)$ transitions of the  ground band of $^{48}$Ti.
The  $B(E2)$ values calculated  with the D171 potential (cal-1)  and 
 the $L$-dependent  $V$ (MeV) (cal-2)  are compared with the experimental data \cite{Ernst2000}. 
}
\begin{center}
\begin{tabular}{rccccc}
 \hline
  \hline
  $J^\pi$   & $<R^2>^{1/2}$&  \multicolumn{2}{c}{$B(E2)$}    
 & $V$ &$B(E2)$\\
    &   cal-1  & cal-1&exp.\cite{Ernst2000}& cal-2 &cal-2\\
   \hline
  $0^+$   
         &  4.41 &     &            &  171.0&  \\ 
  $2^+$  
       &4.37  &13.6 &15.0  &169.4  &  13.8 \\  
  $4^+$   
    	 &  4.33 &17.9&18.4   & 167.7 & 18.8  \\ 
\hline
 \hline
\end{tabular}
\end{center}
\end{table}

\par 
As for the  $N=12$ ground band,   the D181 potential  with   $J_v$=388 MeVfm$^3$ determined at $E_\alpha$=18 MeV locates the ground state $0^+$, which  overbinds  -4.27  MeV, compared with the experimental ground state. 
When discussing the ground band deep below the threshold, as was the case  in $^{52}$Ti \cite{Ohkubo2020}, the potential  must be readjusted by  taking  into account that  the   volume integral of  the real potential decreases   toward the threshold due to the threshold anomaly \cite{Mahaux1986}.  
With  a decreased potential strength, $V$=171 MeV (D171)  with $J_v$=367 MeVfm$^3$,  the calculated ground band in Fig.~5(b) falls well in  correspondence with the experimental   ground band in Fig.~5(c).  
   In Table II  calculated rms intercluster radii and  $B(E2)$ values  are given. The $B(E2)$ values  are  calculated  with the D171 potential and  the  $L$-dependent  $V$   tuned to reproduce the experimental excitation energy of the ground band 
because     the  $L$-dependence of  $V$ has been known widely, for example,
  in    $^{20}$Ne \cite{Michel1989}, $^{44}$Ti \cite{Michel1986,Michel1988}, 
   $^{94}$Mo \cite{Ohkubo1995,Souza2015}, $^{212}$Po \cite{Ohkubo1995,Ni2011}and $^{46,50}$Cr  \cite{Mohr2017}. 
    A small effective charge $\Delta e=0.1 e$ is introduced for protons and neutrons.
The experimental $B(E2)$ values \cite{Ernst2000} are reproduced well.
This small effective charge seems reasonable considering that core excitations are important in the  $^{44}$Ti region \cite{Arnswalda2017,Ohkubo1998B} and that
  the observed     $B(E2:2_1^+ (1.157 {\rm MeV}) \rightarrow {\rm g.s.})$ of  $^{44}$Ca,    10.9 W.u.,  is large.  
The rms charge radius  $<r^2>^{1/2}$=3.71 fm of the ground state
calculated using  the experimental values  $<r^2>_{\alpha}^{1/2}$=1.676 fm  and $<r^2>_{\rm ^{44}Ca}^{1/2}$=3.518 fm  \cite{Angeli2013}  is to be compared 
 with the  experimental value 3.59 fm\cite{Angeli2013}. The calculated  intercluster distance of the ground state is about 85\% of the sum of the experimental rms charge radii of the two clusters, which is only slightly small compared to  87\% for the ground state of $^{44}$Ti \cite{Michel1988}. 
For $J=0^+$, the overlaps of the six  deeply bound forbidden states supported by the potential with the harmonic oscillator wave functions  with 
$\hbar \omega$=10.41 MeV, which is nearly equal to   $\hbar \omega$=10.5 MeV used in the {\it ab initio} calculations of $^{48}$Ti in Ref. \cite{Vary2009}, are
1.000, 1.000, 0.999, 0.999, 0.996 and  0.977 for the oscillator quanta $N_{HO}$=$0$, $2$, $\cdots$, $10$, respectively.  This means that the obtained ground state wave function is orthogonal to the Pauli-forbidden states with $N_{HO}\leq10$ in the resonating group method  mimicking   Saito's orthogonality condition model \cite{Saito1968}.  The   probability with quanta $N_{HO}$ component in the ground state wave function is  78.0, 11.3, 4.2, 1.8 and 1.0 \% for    $N_{HO}=12, 14,$  $\cdots$, $20$, respectively.   The dominant  $N_{HO}=12$  shell-model like component 78\% is larger  than that of $^{44}$Ti in Ref. \cite{Michel1988}.  The existence of  the  significant amount of higher $N_{HO}\geq14$ components, however,  means that  the ground state  has   $\alpha$ clustering if  modest. The $2^+$ and $4^+$ states have  similar characters. This means that  $\alpha$ clustering, i.e., sizable four-particle excitations from the $fp$-shell to the higher major shells, contributes not only to the large $B(E2)$ values but also to
  the  $0\nu\beta\beta$ decay half-life   of $^{48}$Ca,
   performing  longer than the  shell model within the $N_{HO}=12$ model space.

\section{DISCUSSION AND SUMMARY}
 %%% Fig.6  alpha-44Ca pot Wintgen  Luneburg lens pot 
 \begin{figure}[t]
\includegraphics[width=7.0cm]{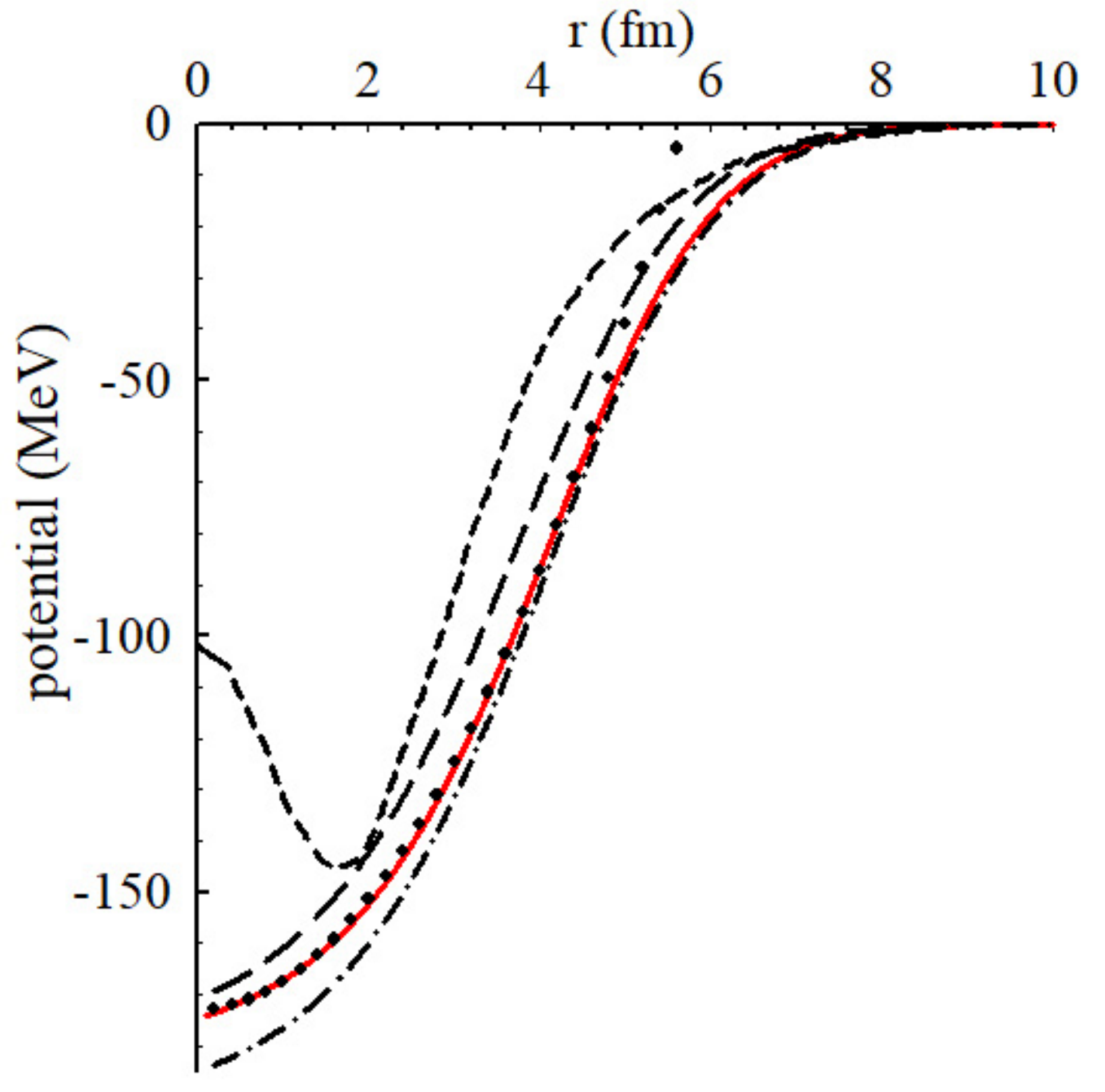}% Here is how 
 \protect\caption{(Color online) The D181 $\alpha$+$^{44}$Ca potential at $E_\alpha=$18 MeV  (solid line) is  compared with the Luneburg lens potential \cite{Michel2002,Ohkubo2016,Luneburg1964} with   $R_0$=5.68 fm   and $V_0$=173 MeV (points) and     the energy-independent equivalent local potential 
  for $L=$0 in  the GCM calculation    \cite{Wintgen1983} (medium dashed line).
  For comparison   the  $\alpha$+$^{40}$Ca  potential used   in the $\alpha$ cluster calculation of $^{44}$Ti       \cite{Michel1988}  (long dashed line) and  
    the      $\alpha$+$^{48}$Ca    potential     at $E_\alpha=$18 MeV
      \cite{Ohkubo2020} (dashed-dot line) are also displayed.
      }
 \end{figure}
 
\par
In Fig.~6 the  D181 potential  is compared    with a Luneburg lens \cite{Luneburg1964} potential, which  is  a truncated harmonic oscillator potential \cite{Michel2002,Ohkubo2016} given  by 
$V(r) = V_0 \left(  {r^2}/{R_0^2}-1 \right)$ for $r \leq R_0$ and  $V(r) = 0$ for 
$r > R_0$.
The potential  resembles the Luneburg lens potential in the internal region,
  which explains   why the   potential embeds the deeply bound  Pauli-forbidden states with $N<$12  and locates the Pauli-allowed $N=12$  band  in correspondence to the  experimental ground band.
The energy dependence of   $J_v$=388 MeVfm$^3$ at  18 MeV and 
  356 MeVfm$^3$ at  24.1 MeV are consistently  in line with      $J_v$=340 MeVfm$^3$    at $E_\alpha$=29  MeV in Ref.  \cite{Michel1979}.
 The D181 potential in Fig.~6 is reasonable being in-between the potentials for   $\alpha$+$^{40}$Ca and $\alpha$+$^{48}$Ca.
  The equivalent local potential  of the  microscopic GCM cluster model calculation\cite{Langanke1982,Wintgen1983}    belongs to a shallower potential family, 
which   explains why     the     $N=$12 band       appears above the $\alpha$ threshold energy         and there appears no state  corresponding to the ground state of $^{48}$Ti. 

%\section{SUMMARY}
\par
To summarize, 
the $\alpha$+$^{44}$Ca fusion excitation function at $E_\alpha$=9-18  was reproduced well by a global Luneburg lens-like potential, which fits $\alpha$+$^{44}$Ca scattering   at $E_\alpha$=18-100 MeV.
The existence of  a $7^-$  state  with the $\alpha$+$^{44}$Ca cluster structure of the higher nodal $N=$15 band in  $^{48}$Ti was confirmed for the first time at  $E_x=18.8$ MeV in the bump of the   fusion excitation function. 
   The first Pauli-allowed     $N=$12 band with the  $\alpha$+$^{44}$Ca cluster structure     falls below the $\alpha$ threshold in correspondence  well with the experimental ground band of $^{48}$Ti. The   experimental $B(E2)$ values
of the ground band   were reproduced well by  calculations and  the enhancement is found to be due to $\alpha$ clustering.


\begin{thebibliography}{reference}
%%%%%%%%%%%
\bibitem {Suppl1972}%Suppl 1972
 K. Ikeda {\it et al.}, 
%  {\it Alpha-like four-body correlations molecular aspects in nuclei}
% (Publication Office, Progress of Theoretical Physics Supplement No. 52, 1972).
Prog. Theor. Phys.   Suppl. No. {\bf 52},  1 (1972) and references therein.
\bibitem {Suppl1980}%Suppl 1980
 K. Ikeda {\it et al.}, 
%  {\it Comprehensive study of structure of light nuclei}
% (Publication Office, Progress of Theoretical Physics Supplement No. 68, 1980).
 Prog. Theor. Phys.   Suppl. No. {\bf 68},  1 (1980) and references therein.\bibitem{Ohkubo1998}
 S. Ohkubo, M. Fujiwara, and P. E. Hodgson, 
 Prog. Theor. Phys. Suppl. {\bf 132}, 1 (1998) and references therein. 
% \bibitem {Ohkubo1998A}
% S. Ohkubo, {\it Alpha-clustering and molecular structure of medium-weight
%and heavy nuclei} (Publication Office, Progress of Theoretical
%Physics Supplement No. 132, 1998).
\bibitem {Ohkubo1999}
S. Ohkubo, T. Yamaya, and P. E. Hodgson,
Nuclear clusters, in {\it Nucleon-Hadron Many-Body
Systems}, (edited by H. Ejiri and H. Toki) 
(Oxford University Press, Oxford,1999), p. 150 
and references therein.
 \bibitem {Michel1986}
F. Michel, G. Reidemeister, and S. Ohkubo, 
Phys. Rev. Lett. {\bf 57}, 1215 (1986).
\bibitem {Michel1988}
 F. Michel, G. Reidemeister, and S. Ohkubo,
 Phys. Rev. C {\bf 37}, 292 (1988).
 \bibitem {Ohkubo1988} %alpha+40Ca HNY effective interaction
 S. Ohkubo, Phys. Rev. C {\bf 38},  2377 (1988).
 \bibitem {Yamaya1990B} 
T. Yamaya, S. Oh-ami, M. Fujiwara, T. Itahashi, K. Katori, M. Tosaki, S. Kato, S. Hatori, and S. Ohkubo, Phys. Rev.  C {\bf 42},1935 (1990).
\bibitem {Guazzoni1993}
P. Guazzoni, M. Jaskola, L. Zetta, C. Y. Kim, T. Udagawa, and
G. Bohlen, Nucl. Phys. {\bf A564}, 425 (1993).

 \bibitem{Yamaya1998}
T. Yamaya, K. Katori, M. Fujiwara, S. Kato, and S. Ohkubo,
Prog. Theor. Phys. Suppl. {\bf 132}, 73 (1998).
\bibitem{Sakuda1998}
 T. Sakuda and  S. Ohkubo, Prog. Theor. Phys. Suppl. {\bf 132}, 103
(1998).
\bibitem{Ohkubo1998B}%alpha-cluster structure of 44Ti in core-excited α+40Ca model,
S. Ohkubo, Y. Hirabayashi, and T. Sakuda,
 Phys. Rev. C {\bf 57}, 2760 (1998).
\bibitem{Michel1998}
 F. Michel, S. Ohkubo, and G. Reidemeister, Prog. Theor. Phys.
Suppl. {\bf 132}, 7 (1998).
\bibitem{Fukada2009}% 44Ti 7-;  46Ti many states observed, 52Ti exp
M. Fukada, M. K. Takimoto, K. Ogino, and S. Ohkubo, Phys.
Rev. C {\bf 80}, 064613 (2009). 

\bibitem {Ohkubo2020} %52Ti=alpha+48Ca
S. Ohkubo,
Phys. Rev. C  {\bf 101},  041301(R) (2020).

%48Ti
 \bibitem {Anantaraman1975} %(6Li,d) alpha transfer
 N. Anantaraman  {\it et al.}, Phys. Rev. Lett. {\bf 35}, 1131 (1975).
 \bibitem {Fulbright1977}  %44Ca(6Li,d) alpha transfer  48Ti
H. W. Fulbright, C. L. Bennett, R. A. Lindgren, R. G. Markham, S. C. McGuire, G. C. Morrison, U.  Strohbusch,  and J. T\={o}ke,
Nucl. Phys.  {\bf A284}, 329 (1977).

 \bibitem {Carey1981} %(p,p alpha) knockout reaction  48Ti(p,p alpha) 44Ca 
T. A. Carey,  P. G. Roos, N. S. Chant, A. Nadasen, ~ and H. L. Chen,
Phys. Rev. C  {\bf 23},  576 (1981).
\bibitem {Carey1984} %(p,p alpha) knockout reaction  48Ti(p,p alpha) 44Ca 
T. A. Carey,  P. G. Roos, N. S. Chant, A. Nadasen, ~ and H. L. Chen,
Phys. Rev. C  {\bf 29}, 1273 (1984).
\bibitem {Roos1984} %clustering aspects of medium and high energy reactions with light ions  Chester conference (p,p alpha)
  P. G. Roos, {\it Clustering Aspects of Nuclear Structure},  (edited by J. S. Lilley and M. A. Nagarajan)  (D. Reidel Publishing Company, Dordrecht, 1985), p.279.
%  \bibitem {Taniguchi2021} 
%Y. Taniguchi, K. Yoshida, Y. Chiba, Y. Kanada-Enyo, M. Kimura, and K. Ogata,
%Phys. Rev. C  {\bf 103}, L031305 (2021).

\bibitem{Oertzen2001}%Covalently bound molecular structures in the alpha+16O system
W. von Oertzen, Eur. Phys. J. A {\bf 11}, 403 (2001). 

% double beta decay 
\bibitem{Tetsuno2020}%Status of 48Ca double beta decay search and its future prospect in CANDLES, 
K. Tetsuno, S. Ajimura, K. Akutagawa, T. Batpurev {\it et al.},
%W. M. Chan, K. Fushimi, R. Hazama, T. Iida,　Y. Ikeyama, B. T. Khai, T. Kishimoto, K. K. Lee, X. Li,　K. Matsuoka, K. Matsuoka, K. Mizukoshi, Y. Mori,　K. Nakajima, P. Noithong, M. Nomachi, I. Ogawa, H. Ohsumi, K. Ozawa, K. Shimizu, M. Shokati, F. Soberi, K. Suzuki, Y. Takemoto, Y. Takihira, Y. Tamagawa, M. Tozawa, V. T. T. Trang,  S. Umehara, K. Yamamoto,S. Yoshida, I. Kim, D. H. Kwon, H. L. Kim, H. J. Lee, M. K. Lee, and Y. H. Kim, 
J. Phys. Conf. Series {\bf 1468}, 012132 (2020).
\bibitem{Vergados2012}%Theory of neutrinoless double-beta decay
J. D. Vergados, H. Ejiri, and F. $\breve{\rm S}$imkovic,
Rep. Prog. Phys. {\bf 75}, 106301 (2012).
\bibitem{Engel2017}%Status and future of nuclear matrix elements for neutrinoless double-beta decay:
J. Engel and J. Men\'{e}ndez,
%see for, example,
 Rep. Prog. Phys. {\bf 80}, 046301 (2017).
% shell model 48Ti
\bibitem{Caurier2008} 
E. Caurier, J. Men\'{e}ndez, F. Nowacki, and A. Poves,
Phys. Rev. Lett. {\bf 100}, 052503 (2008).
\bibitem{Horoi2013} 
M. Horoi, Phys. Rev. C {\bf 87}, 014320 (2013).
\bibitem{Iwata2016}%Large-Scale Shell-Model Analysis of the Neutrinoless beta beta  Decay of 48Ca
Y. Iwata, N. Shimizu, T. Otsuka, Y. Utsuno, J. Men\'{e}ndez, M.  Honma, and T. Abe,
Phys. Rev. Lett. {\bf 116}, 112502 (2016).
\bibitem{Coraggio2020}%Calculation of the neutrinoless double- decay matrix  element within the realistic shell model,
L. Coraggio, A. Gargano, N. Itaco, R. Mancino, and
F. Nowacki,  Phys. Rev. C {\bf 101}, 044315 (2020).
\bibitem{Yao2020} %Ab initio treatment of collective correlations and the neutrinoless double beta decay of 48Ca,
J. M. Yao, B. Bally, J. Engel, R. Wirth, T. R. Rodr\'{i}guez,
and H. Hergert,
Phys. Rev. Lett. {\bf 124}, 232501 (2020).
\bibitem{Belley2021}%ab Initio Neutrinoless Double-Beta Decay Matrix Elements for 48Ca, 76Ge, and 82Se
A. Belley, C. G. Payne, S. R. Stroberg, T. Miyagi, and J. D. Holt,
Phys. Rev. Lett. {\bf 126}, 042502 (2021).
% QRPA
\bibitem{Simkovic2013}
F. $\breve{\rm S}$imkovic, V. Rodin, A. Faessler, and P. Vogel, Phys. Rev.
C {\bf 87}, 045501 (2013).
 \bibitem{Hyvarinen2015}
 J. Hyv$\ddot{\rm a}$rinen and J. Suhonen, Phys. Rev. C {\bf 91}, 054308 (2015).
\bibitem{Terasaki2015}
 J. Terasaki, Phys. Rev. C {\bf 91}, 034318 (2015).

% Projected HF 
\bibitem{Rath2010} P. K. Rath, R. Chandra, K. Chaturvedi, P. K. Raina, and
J. G. Hirsch, Phys. Rev. C {\bf 82}, 064310 (2010).
% GCM %48Ti
\bibitem{Rodriguez2010} 
T. R. Rodriguez and G. Martinez-Pinedo, Phys. Rev.
Lett. {\bf 105}, 252503 (2010).
\bibitem{Hinohara2014} 
N. Hinohara and J. Engel,
 Phys. Rev. C {\bf 90}, 031301(R) (2014).
%density functional
\bibitem{Vaquero2013}
 N. L. Vaquero, T. R. Rodr\'{i}guez, and J. L. Egido, Phys. Rev.
Lett. {\bf 111}, 142501 (2013).
\bibitem{Yao2015}
 J. M. Yao, L. S. Song, K. Hagino, P. Ring, and J. Meng,
Phys. Rev. C {\bf 91}, 024316 (2015).
%IBM
\bibitem{Barea2012} 
J. Barea and F. Iachello, Phys. Rev. C {\bf 79}, 044301 (2009);
J. Barea, J. Kotila, and F. Iachello, Phys. Rev. Lett. {\bf 109},
042501 (2012).
% alpha+44Ca GCM 
\bibitem {Langanke1982}% alpha+44Ca GCM band structure 
 K. Langanke, Nucl. Phys. {\bf A377}, 53 (1982).
 \bibitem {Wintgen1983}% local potential for alpha+44Ca  
 D. Wintgen, H. Friedrich, and K. Langanke, Nucl, Phys.  {\bf A408}, 239 (1983).

\bibitem{Bailey2021}% 52Ti wavelet 
S. Bailey,% {\it et al.}, 
T. Kokalova, M. Freer, C. Wheldon, R. Smith, J. Walshe
 {\it et al.},
% N. Soi\'{c}, L. Prepolec, V. Toki\'{c},  L. Prepolec, F. M. Marqu\'{e}s, L. Achouri, F. Delaunay, D. M. Parlog,  Q. Deshayes, B. Fern\'{a}ndez-Dominguez,  and B. Jacquot, 
Eur. Phys. J. A {\bf 57}, 108 (2021). 
 \bibitem{Bailey2019}%%Extracting the spectral signature of alpha clustering in 44Ti,48Ti,52Ti using a continuous wavelet transform 
S. Bailey,% {\it et al.},  
T. Kokalova, M. Freer, C. Wheldon, R. Smith, J. Walshe  {\it et al.},
% N. Curtis, N. Soi\'{c}, L. Prepolec, V. Toki\'{c}, F. M. Marqu\'{e}s, L. Achouri, F. Delaunay,  Q. Deshayes, D. M. Parlog, B. Fern\'{a}ndez-Dominguez, B. Jacquot, and A. Soylu, 
Phys. Rev. C {\bf 100}, 051302(R) (2019).
\bibitem{Michel2002}%Luneburg lens 
F. Michel, G. Reidemeister, and S. Ohkubo,
 Phys. Rev. Lett. {\bf 89}, 152701 (2002).
 \bibitem{Ohkubo2016}%Luneburg lens-like potential
S. Ohkubo, Phys. Rev. C  {\bf 93}, 041303(R) (2016).  
 \bibitem {Luneburg1964} %Luneburg lens 
  R. K. Luneburg, {\it Mathematical Theory of Optics} (University of California Press, Oakland, California, 1964).
\bibitem {Budzanowski1966}
A. Budzanowski, K. Grotowski, L. Jarczyk, B. {L}azarka, S. Micik, H. Niewodnicza\'{n}ski, A. Strza{l}kowski, and Mrs. Z. Wr\'{o}bel, 
Phys. Lett. {\bf 16}, 135 (1965). 
\bibitem {Gruhn1966}
C. R. Gruhn and N. S. Wall,
 Nucl. Phys. {\bf 81}, 161 (1966). 
 
\bibitem {Gaul1969}
 G.  Gaul, H. L$\ddot{\rm u}$decke, R. Santo, H. Schmeing, and R. Stock,
 NucI. Phys.  {\bf A137}, 177 (1969).

 \bibitem{Delbar1978} %alpha+44Ca pot
 Th. Delbar,
 Gh. Gr\'{e}goire, G. Pai\'{c}, R. Ceuleneer, 
F. Michel, R. Vanderpoorten  {\it et al.},
%A. Budzanowski, H. Dabrowski, L. Freindl, 
%K. Grotowski, S. Micek, R. P\l aneta, A. Strzalkowski, and K. A. Eberhard,
 Phys. Rev. C  {\bf 18}, 1237 (1978).

\bibitem{Albinski1982} % alpha+44Ca E=29 MeV B/I decomposition fig one order small but internal wave survives
  J. Albi\'{n}ski and F. Michel,
 Phys. Rev. C  {\bf 25}, 213  (1982). 
 \bibitem {Brink1985}
 D. M. Brink, {\it Semi-Classical Methods for Nucleus-Nucleus Scattering}
(Cambridge University Press, Cambridge,U.K., 1985).
 \bibitem {Michel1986A} %molecular interpretation fusion 
F. Michel, G. Reidemeister, and S. Ohkubo, 
Phys. Rev. C  {\bf 34}, 1248 (1986).
  \bibitem {Eberhard1979} % fusion alpha+44Ca=48Ti
K. A. Eberhard, C. Appel, R. Bangert, L. Cleemann, J. Eberth, and V. Zobel,
Phys. Rev. Lett. {\bf 43}, 107 (1979).
 
 %fusion
\bibitem{Ohkubo1987A} %alpha+40Ca fusion
S. Ohkubo and D. M. Brink,
Phys. Rev. C  {\bf 36},  966 (1987).
\bibitem{Ohkubo1987B} %12C+12C fusion
S. Ohkubo and D. M. Brink,
Phys. Rev. C  {\bf 36},  1375 (1987).
 \bibitem{Mosel1984}
 U. Mosel,  in {\it Treatise on Heavy Ion Science}, edited by D. A. Bromley,  (Plenum, New York, 1984), p.3 and see references therein.
\bibitem {Mahaux1986}
 C. Mahaux, H. Ngo, and G. R. Satchler, 
 Nucl. Phys.  {\bf A449}, 354 (1986); 
  C. Mahaux, H. Ngo, and G. R. Satchler, 
 {\bf A456}, 134 (1986). % dispersion alpha+16O alpha+40Ca
\bibitem {Swan1955}
P. Swan,
Proc . R. Soc. {\bf A  228}, 10 (1955). 

   \bibitem{Beck2020}%Status on  fusion at deep sub barrier energies: impact of resonances on astrophysical  factors
 C. Beck, A. M. Mukhamedzhanov, and X. Tang,
Eur. Phys. J.  A {\bf 56}, 87 (2020). 
 \bibitem{Jiang2021}%Heavy-ion fusion reactions at extreme sub-barrier energies
 C. L. Jiang, B. B. Back, K. E. Rehm, K. Hagino, G. Montagnoli, and  A. M. Stefanini,
Eur. Phys. J. A  {\bf 57}, 235 (2021) and references therein.

\bibitem {Nemoto1972}
F. Nemoto and H. Bando,  
 Prog. Theor. Phys.  {\bf 47}, 1210 (1972).
\bibitem {Hiura1972}%suppl 52
J. Hiura, F. Nemoto, and H. Bando, 
 Prog. Theor. Phys. Suppl. {\bf 52}, 173 (1972). 
\bibitem {Ohkubo1977}
S. Ohkubo, Y. Kondo, and S. Nagata, 
Prog. Theor. Phys.  {\bf 57}, 82 (1977). 
\bibitem {Fujiwara1980}
 Y. Fujiwara, H. Horiuchi, K. Ikeda, M. Kamimura, K. Kato, Y. Suzuki, and E. Uegaki,
Prog. Theor. Phys. Suppl.  {\bf 68}, 29 (1980) and references therein.
 \bibitem{Burrows2006}%Nuclear  Data  Sheets  for  A  =  48
T. W. Burrows,
Nucl. Data Sheets {\bf 107}, 1747 (2006). 
% L-dependence
\bibitem {Michel1989}
F. Michel, Y. Kondo, and G. Reidemeiter,
Phys. Lett. {\bf B 220}, 479 (1989).
 \bibitem {Ohkubo1995}% alpha+90Zr , alpha+208Pb
S. Ohkubo, Phys. Rev. Lett.   {\bf 74}, 2176 (1995).
\bibitem{Souza2015} %α-cluster structure in even-even nuclei around 94Mo
M. A. Souza and H. Miyake,
Phys. Rev. C  {\bf 91}, 034320 (2015).
\bibitem{Ni2011}%alpha-cluster structure above doubly closed shells in a generalized density-dependent cluster model
D. Ni and Z. Ren,
Phys. Rev. C  {\bf 83}, 014310 (2011).
\bibitem{Mohr2017}
P. Mohr, % 46,54Cr double folding model
Eur. Phys. J. A  {\bf 53}, 209 (2017).


  \bibitem{Ernst2000}%Stringent Tests of Shell Model Calculations in fp Shell Nuclei 46,48Ti and 50,52Cr from Measurements of g Factors and BE2 Values
R. Ernst {\it et al.}, 
%, K.-H. Speidel, O. Kenn, U. Nachum, J. Gerber, P. Maier-Komor, N. Benczer-Koller, G. Jakob, G. Kumbartzki, L. Zamick, and F. Nowacki,
Phys. Rev. Lett.  {\bf 84}, 416  (2000);
 % Shell structure of Ti and Cr nuclei from measurements of g factors and lifetimes
R. Ernst {\it et al.}, 
%, K.-H. Speidel, O. Kenn, A. Gohla, U. Nachum, J. Gerber, P. Maier-Komor, N. Benczer-Koller, G. Kumbartzki, G. Jakob, L. Zamick, and F. Nowacki,
Phys. Rev. C {\bf 62}, 024305 (2000).
 \bibitem{Arnswalda2017}%Enhanced collectivity along the N=Z line: Lifetime measurements in 44Ti, 48Cr, and 52Fe
K. Arnswald {\it et al.}, 
%, T. Braunroth, M. Seidlitz, L. Coraggio, P. Reiter, B. Birkenbach, A. Blazhev, A. Dewald, C. Fransen, B. Fu, A. Gargano, H. Hess, R. Hirsch, N. Itaco, S. M. Lenzi, L. Lewandowski, J. Litzinger, C. M$\ddot{\rm u}$ller-Gatermann, M. Queiser, D. Rosiak, D. Schneiders, B. Siebeck, T. Steinbach, A. Vogt, K. Wolf, K.O.Zell,
Phys. Lett. B {\bf 772}, 599 (2017).
\bibitem{Angeli2013}%Table of experimental nuclear ground state charge radii: An 
I. Angeli and K. P. Marinova,
Atomic Data and Nucl. Data Tables {\bf 99},  69 (2013).
\bibitem{Vary2009}%A no-core shell model for 48Ca, 48Sc and 48Ti
J. P. Vary, S. Popescu, S. Stoica, and P. Navr\'{a}til,
J. Phys. G  Nucl. Part. Phys. {\bf 36}, 085103 (2009).
\bibitem{Saito1968}
S.~Saito, Prog.~Theor.~Phys. {\bf 40},  893 (1968); 
S.~Saito, Prog.~Theor.~Phys. {\bf 41},  705 (1969). 
   \bibitem {Michel1979} %alpha+44Ca  spline  pot
F. Michel and R. Vanderpoorten, 
Phys. Lett. {\bf 82 B}, 183 (1979).
\end{thebibliography}
\end{document}